\documentstyle[12pt]{article}
\catcode`@=11
\newcount\@tempcntc
\def\@citex[#1]#2{\if@filesw\immediate\write\@auxout{\string\citation{#2}}\fi
  \@tempcnta\z@\@tempcntb\m@ne\def\@citea{}\@cite{\@for\@citeb:=#2\do
    {\@ifundefined
       {b@\@citeb}{\@citeo\@tempcntb\m@ne\@citea\def\@citea{,}{\bf ?}\@warning
       {Citation `\@citeb' on page \thepage \space undefined}}%
    {\setbox\z@\hbox{\global\@tempcntc0\csname b@\@citeb\endcsname\relax}%
     \ifnum\@tempcntc=\z@ \@citeo\@tempcntb\m@ne
       \@citea\def\@citea{,}\hbox{\csname b@\@citeb\endcsname}%
     \else
      \advance\@tempcntb\@ne
      \ifnum\@tempcntb=\@tempcntc
      \else\advance\@tempcntb\m@ne\@citeo
      \@tempcnta\@tempcntc\@tempcntb\@tempcntc\fi\fi}}\@citeo}{#1}}
\def\@citeo{\ifnum\@tempcnta>\@tempcntb\else\@citea\def\@citea{,}%
  \ifnum\@tempcnta=\@tempcntb\the\@tempcnta\else
   {\advance\@tempcnta\@ne\ifnum\@tempcnta=\@tempcntb \else \def\@citea{--}\fi
    \advance\@tempcnta\m@ne\the\@tempcnta\@citea\the\@tempcntb}\fi\fi}
\catcode`@=12
\newcommand\jmin{j_{\rm min}}
\newcommand\one{1\kern-2.5pt{\rm l}}
\newcommand\vslash{v\kern-5pt\raise1pt\hbox{$\scriptstyle/$}}
\newcommand\Frac[2]{\hbox{$\frac{#1}{#2}$}}
\begin{document}
\begin{flushright}
MZ-TH/96-xx \\[-0.2cm]
January 1996 \\[-0.2cm]
\end{flushright}
\begin{center}

{\Large\bf Current, Pion and Photon Transitions} \\[.3cm] 
{\Large\bf between Heavy Baryons}\\[1.75cm]

{\large J\"urgen G.~K\"orner\footnote{Supported in part by BMBF, FRG under 
  contract 06MZ730}} \\[.4cm]
Institut f\"ur Physik, Johannes Gutenberg-Universit\"at \\
Staudingerweg 7, D-55099 Mainz, Germany.

\end{center}
\vspace{1.5cm}
\centerline {\bf ABSTRACT}\noindent
I discuss the structure of current-induced bottom baryon to charm baryon 
transitions, and the structure of pion and photon transitions between heavy 
charm or bottom baryons in the Heavy Quark Symmetry limit as
${m_Q\rightarrow\infty}$. The emphasis is on the structural similarity of 
the Heavy Quark Symmetry predictions for the three types of transitions. 
The discussion involves the ground state $s$-wave heavy baryons as well as 
the excited $p$-wave heavy baryon states. Using a constituent quark model 
picture for the light diquark system with an underlying 
$SU(2N_f)\otimes O(3)$ symmetry one arrives at a number of new predictions 
that go beyond the Heavy Quark Symmetry predictions.

\newpage

\section{Introduction}

Because of the initial abundance of data on heavy charm and bottom mesons 
the attention of experimentalists and theoreticians had initially been 
directed towards applications of the Heavy Quark Effective Theory (HQET) 
to the meson sector. In the meantime the situation has considerably changed 
and data on heavy baryons and their decay properties are starting to become 
available in impressive amounts. Since theoretical results on the 
description of semileptonic, one-pion and photon decays of heavy baryons are
widely dispersed in the literature it is worthwhile to review the necessary
theoretical HQET framework to describe these three types of decays in a 
comprehensive manner. The emphasis is on the structural similarity of 
the HQS description of these decays. In this review I am only concerned 
with the leading order contributions to these decays, leading in terms of 
the inverse heavy quark mass expansion provided by HQET. At the end of my 
presentation I discuss possibilities to further constrain the Heavy Quark 
Symmetry (HQS) structure of the respective decays by resolving the light 
diquark transitions in terms of a constituent quark model description of 
the light diquark transitions with an underlying $SU(2N_f)\otimes O(3)$ 
symmetry.

\section{Heavy Baryon Spin Wave Functions}

Let us begin by constructing the heavy baryon spin wave functions that
enter into the descriptions of heavy baryon decays. A heavy baryon is made 
up of a light diquark system $(qq)$ and a heavy quark $Q$. The light 
diquark system has bosonic quantum numbers $j^P$ with total angular 
momentum $j=0,1,2 \dots$ and parity $P=\pm 1$. To each diquark system with 
spin-parity $j^P$ there is a degenerate heavy baryon doublet with 
$J^P=(j\pm\Frac12)^P$ ($j=0$ is an exception). It is important to realize 
that the HQS structure of the heavy baryon states is entirely determinated 
by the spin-parity $j^P$ of the light diquark system. The requisite angular 
momentum coupling factors can be read off from the coupling scheme
\begin{equation}\label{eqn1}
j^P\otimes\Frac12^+\Rightarrow J^P.
\end{equation}
Apart from the angular momentum coupling factors the dynamics
of the light system is completely decoupled from the heavy quarks. 

Let us cast these statements into a covariant framework in which the heavy
baryon wave function $\Psi$ describes the amplitude of finding the light 
diquark system and the heavy quark in the heavy baryon. The covariant 
equivalent of the coupling scheme Eq.~(\ref{eqn1}) is then given by
\begin{equation}\label{eqn2}
\Psi=\hat\phi_{\mu_1\cdots\mu_j}\psi^{\mu_1\cdots\mu_j},
\end{equation}
where $\hat\phi_{\mu_1\cdots\mu_j}$ stands for the tensor representation of 
the spin-parity $j^P$ diquark state and $\psi^{\mu_1\cdots\mu_j}$ represents 
the heavy-side baryon spin wave function (in short: heavy baryon wave 
function) coupling the heavy quark to the heavy baryon. Let us be more 
specific. If
\begin{equation}\label{eqn3}
|J^P,m_J\rangle=\sum_{m_j+m_Q=m_J}\!\!\!\!\!\!
  \langle j^P,m_j;\Frac12^+,m_Q|J^P,m_J\rangle
  |j^P,m_j\rangle|\Frac12^+,m_Q\rangle
\end{equation}
defines the static quark model wave function, the C.G. coefficients
determining the heavy quark - light diquark content of the heavy
baryon can be obtained in covariant fashion from the heavy baryon
spin wave function by the covariant projection
\begin{equation}\label{eqn4}
\langle j^P,m_j;\Frac12^+,m_Q|J^P,m_J\rangle
  =\varepsilon^*_{\mu_1\cdots\mu_j}(m_j)\bar u(m_Q)
  \psi^{\mu_1\cdots\mu_j}(m_J).
\end{equation}
The r.h.s. of Eq.~(\ref{eqn4}) can be evaluated for any velocity $v_\mu$ of 
the heavy baryon which, at leading order, equals the velocity of the heavy 
quark and the diquark system. Details including questions of normalization 
can be found in~\cite{strasba1,strasba5}. Differing from~\cite{strasba1} we 
have normalized the spinors appearing in Eq.~(\ref{eqn4}) to $1$ and not to 
$2M$ and $2M_Q$ as in~\cite{strasba1}. It is not difficult to construct the 
appropiate heavy baryon spin wave functions using the heavy quark on-shell 
constraint $\vslash\psi^{\mu_1\cdots\mu_j}=\psi^{\mu_1\cdots\mu_j}$ and the 
appropiate normalization condition. In Table 1 (fourth column) we have 
listed a set of correctly normalized heavy baryon spin wave functions that 
are associated with the diquark states $j^P=0^+,1^+,0^-,1^-,2^-$.

Next we turn our attention to the question of which low-lying heavy
baryon states can be expected to exist. From our experience with light 
baryons and light mesons we know that one can get a reasonable description 
of the light particle spectrum in the constituent quark model picture. This 
is particularly true for the enumeration of states, their spins and their 
parities. As much as we know up to now, gluon degrees of freedom do not 
seem to contribute to the particle spectrum. It is thus quite natural to 
try the same constituent approach to enumerate the light diquark states, 
their spins and their parities. 

From the spin degrees of freedom of the two light quarks one obtains a 
spin~0 and a spin~1 state. The total orbital state of the diquark system 
is characterized by two angular degrees of freedom which we take to be the 
two independent relative momenta $k=\frac12(p_1-p_2)$ and 
$K=\frac12(p_1+p_2-2p_3)$ that can be formed from the two light 
quark momenta $p_1$ and $p_2$ and the heavy quark momentum $p_3$. 
The $k$-orbital momentum describes relative orbital excitations of the two 
quarks, and the $K$-orbital momentum describes orbital excitations of the 
center of mass of the two light quarks relative to the heavy quark. The 
$(k,K)$-basis is quite convenient in as much as it allows one to classify 
the diquark states in terms of $SU(2N_f)\otimes O(3)$ representations as 
will be discussed later on. Table~1 lists all ground state $s$-wave and 
excited $p$-wave heavy baryon wave functions as they occur in the constituent 
approach to the light diquark excitations. They are grouped together in terms 
of $SU(2N_f)\otimes O(3)$ representations with $N_f=2(u,d)$. The $s$-wave 
states are in the $\underline{10}\otimes\underline{1}$ representation, and 
the $p$-wave states are in the $\underline{10}\otimes\underline{3}$ and 
$\underline{6}\otimes\underline{3}$ representation of $SU(4)\otimes O(3)$ 
for the $K$- and $k$-multiplets, respectively. Apart from the ground state 
$s$-wave baryons one thus has altogether seven $\Lambda$-type $p$-wave 
states and seven $\Sigma$-type $p$-wave states. The analysis can easily be 
extended to the case $SU(6)\otimes 0(3)$ bringing in the strangeness quark 
in addition.

Let us mention that, in the charm sector the states $\Lambda_c(2285)$ and
$\Sigma_c(2453)$ are well established while there is first evidence for
the $\Sigma_c^*(2510)$ state. Two excited states $\Lambda_c^{**}(2593)$ 
and $\Lambda_c^{**}(2627)$ have been seen which very likely correspond to 
the two $p$-wave states making up the $\Lambda_{cK1}^{**}$ HQS doublet. 
The charm-strangeness states $\Xi_c(2470)$ and $\Omega_c(2720)$ as well as 
the $\Xi_c^*(2643)$ have been seen. First evidence was presented for the 
$J^P=\frac{1}{2}^+$ state $\Xi'_c(2570)$  with the flavour configuration 
$c\{sq\}$. Thus almost all ground state charm baryons have been seen 
including two $p$-wave states. In the bottom sector the $\Lambda_b(5640)$
has been identified as well as the $\Sigma_b(5713)$ and the 
$\Sigma_b^*(5869)$. Some indirect evidence has been presented for the 
$\Xi_b(5800)$.

\section{Generic Picture of Current, Pion\\ and Photon Transitions}

In Fig.~1 we have drawn the generic diagrams that describe $b\rightarrow c$ 
current transitions, and $c\rightarrow c$ pion and photon transitions 
between heavy baryons in the HQS limit. The heavy-side and light-side 
transitions occur completely independent of each other (they ``factorize'') 
except for the requirement that the heavy side and the light side have the 
same velocity in the initial and final state, respectively, which are also 
the velocities of the initial and final heavy baryons. The $b\rightarrow c$ 
current transition induced by the flavour-spinor matrix~$\Gamma$ is hard 
and accordingly there is a change of velocities $v_1\rightarrow v_2$, 
whereas there is no velocity change in the pion and photon transitions. 
The heavy-side transitions are completely specified whereas the light-side 
transitions $j_1^{P_1}\rightarrow j_2^{P_2}$,
$j_1^{P_1}\rightarrow j_2^{P_2}+\pi$ and
$j_1^{P_1}\rightarrow j_2^{P_2}+\gamma$ are described by a number of form
factors or coupling factors which parametrize the light-side transitions.
The pion and the photon couple only to the light side. In the case of the
pion this is due to its flavour content. In the case of the photon the
coupling of the photon to the heavy side involves a spin flip which is down
by $1/m_Q$ and thus the photon couples only to the light side in the Heavy
Quark Symmetry limit.

Referring to Fig.~1 we are now in the position to write down the generic
expressions for the current, pion and photon transitions according to the
spin-flavour flow depicted in Fig.~1. One has ($\omega=v_1\cdot v_2$)
\medskip\\
{\it current transitions:}
\begin{equation}\label{eqn5}
\bar\psi_2^{\nu_1\cdots\nu_{j_2}}\Gamma\psi_1^{\mu_1\cdots\mu_{j_1}}
\left(\sum_{i=1}^Nf_i(\omega)
t^i_{\nu_1\cdots\nu_{j_2};\mu_1\cdots\mu_{j_1}}\right)
\end{equation}
\begin{eqnarray}
n_1\cdot n_2&=&1\qquad N=\jmin+1\nonumber\\
n_1\cdot n_2&=&-1\quad N=\jmin\nonumber
\end{eqnarray}
{\it pion transitions:}
\begin{equation}\label{eqn6}
\bar\psi_2^{\nu_1\cdots\nu_{j_2}}\psi_1^{\mu_1\cdots\mu_{j_1}}
\left(\sum_{i=1}^Nf_i^\pi
t^i_{\nu_1\cdots\nu_{j_2};\mu_1\cdots\mu_{j_1}}\right)
\end{equation}
\begin{eqnarray}
n_1\cdot n_2&=&1\qquad N=\jmin\nonumber\\
n_1\cdot n_2&=&-1\quad N=\jmin+1\nonumber
\end{eqnarray}
{\it photon transitions:}
\begin{equation}\label{eqn7}
\bar\psi_2^{\nu_1\cdots\nu_{j_2}}\psi_1^{\mu_1\cdots\mu_{j_1}}
\left(\sum_{i=1}^Nf_i^\gamma
t^i_{\nu_1\cdots\nu_{j_2};\mu_1\cdots\mu_{j_1}}\right)
\end{equation}
\begin{eqnarray}
j_1&=&j_2\qquad N=2j_1\nonumber\\
j_1&\neq&j_2\quad N=2\jmin+1\nonumber
\end{eqnarray}
where the $\psi^{\mu_1\cdots\mu_j}$ are the heavy baryon spin wave
functions introduced in Sec.~2.

In each of the above cases we have also given the result of counting the
number~$N$ of independent form factors or coupling factors. These are easy
to count by using either helicity amplitude counting or $LS$ partial wave
amplitude counting. In the case of current and pion transitions the
counting involves the normalities of the light-side diquarks which is
defined by $n=(-1)^jP$.

In the case of the current transitions the tensors 
$t^i_{\nu_1\cdots\nu_{j_2};\mu_1\cdots\mu_{j_1}}$ appearing in 
Eq.~(\ref{eqn5}) have to be build from the vectors $v_1^{\nu_i}$ and 
$v_2^{\mu_i}$, the metric tensors $g_{\mu_i\nu_k}$ and, depending on parity, 
from the Levi-Civita object $\varepsilon(\mu_i\nu_kv_1v_2)
:=\varepsilon_{\mu_i\nu_k\alpha\beta}v_1^\alpha v_2^\beta$. For the pion 
transitions in Eq.~(\ref{eqn6}) the $(j_1+j_2)$-rank tensors 
$t^i_{\nu_1\cdots\nu_{j_2};\mu_1\cdots\mu_{j_1}}$ describing the light-side 
transitions $j_1^{P_1}\rightarrow j_2^{P_2}+\pi$ have to be composed from 
the building blocks $g_{\perp\mu\nu}=g_{\mu\nu}-v_\mu v_\nu$, the pion 
momentum $p_{\perp\mu}=p_\mu-p\cdot v\,v_\mu$ and, depending on parity, from 
the Levi-Civita tensor $\varepsilon(\mu_i\nu_kp\,v)$. Finally, in the photon 
transition case ${j_1}^{P_1}\rightarrow{j_2}^{P_2}+\gamma$ Eq.~(\ref{eqn7}) 
one has to use the field strength tensor 
$F_{\alpha\beta}=k_\alpha\varepsilon_\beta-k_\beta\varepsilon_\alpha$ or,
depending on parity, its dual
$\tilde F_{\alpha\beta}=\frac12\varepsilon_{\alpha\beta\gamma\delta}
F^{\gamma\delta}$ in order to guarantee a gauge invariant coupling of the
photon to the light side. As in the current and pion transition case
further building blocks for the diquark transition tensor are the metric 
tensor, the velocity $v_\alpha$ and the photon momentum $k_\mu$. The number 
of independent tensors that can be written down in each of the three cases 
is necessarily identical to the numbers listed in Eqs.~(\ref{eqn5}), 
(\ref{eqn6}) and~(\ref{eqn7}). Lack of space prevents us from giving the 
explicit forms of these tensors. They can be found in~\cite{strasba1}.

The generic expressions Eq.~(\ref{eqn5}), Eq.~(\ref{eqn6}) and 
Eq.~(\ref{eqn7}) completely determine the HQS structure of the current, 
pion and photon transition amplitudes. It is not difficult to work out 
relations between rates, angular decay distributions etc. from these 
expressions.

It is well worth mentioning that all three covariant coupling expressions
can also be written down in terms of Wigner's 6-$j$ symbol 
calculus~\cite{strasba1,strasba2} as can be appreciated from the discussion 
in Sec.~2 (see Eqs.~(\ref{eqn2}) and~(\ref{eqn3})). For example, looking 
at the pion transition in Fig.~1 one sees that one has to perform 
altogether three angular couplings. They are
\begin{eqnarray}
\mbox{(i)}&&\hspace{1.7cm}{j_1}^{P_1}\otimes\Frac12^+\Rightarrow {J_1}^{P_1}
  \nonumber\\
\mbox{(ii)}&&\hspace{1.7cm}{j_2}^{P_2}\otimes\Frac12^+\Rightarrow {J_2}^{P_2}\\
\mbox{(iii)}&&\hspace{1.7cm}{J_2}^{P_2}\otimes L_\pi\Rightarrow {J_1}^{P_1}
  \nonumber
\end{eqnarray}
where $L_\pi=l_\pi$ is the orbital momentum of the pion and ${J_1}^{P_1}$
and ${J_2}^{P_2}$ denote the $J^P$ quantum numbers of the initial and final
baryons. This is a coupling problem well-known from atomic and nuclear 
physics and the problem is solved by Wigner's 6-$j$ symbol calculus. One 
finds~\cite{strasba1,strasba2}
\begin{eqnarray}
\lefteqn{M^{\pi}(J_1J_1^z\rightarrow J_2J_2^z+L_{\pi}m)}\label{eqn8}\\
  &=&M_{L_\pi}(-1)^{L_\pi+j_2+\Frac12+J}(2j_1+1)^{1/2}(2J_2+1)^{1/2}
  \nonumber\\&&
  \qquad\left\{\begin{array}{ccc}j_2&j_1&L_\pi\\J&J_2&\Frac12\end{array} 
  \right\}\langle LmJ_2J_2^z|J_1J_1^z\rangle,\nonumber
\end{eqnarray}
where the expression in curly brackets is Wigner's 6-$j$ symbol and 
$\langle L_\pi MJ_2J_2^z|J_1J_1^z \rangle$ is the Clebsch-Gordan coefficient 
coupling $L_\pi$ and $J_2$ to $J_1$. $M_{L_\pi}$ is the reduced amplitude of 
the one-pion transition. It is proportional to the invariant coupling 
$f_{l_\pi}$ occurring in the covariant expansion in Eq.~(\ref{eqn6}). 

Let us, for example, calculate the doublet to doublet transition rates for
e.g. $\{\Lambda_{Qk2}^{**}\}\rightarrow\{\Sigma_Q\}+\pi$. The rates are in
the ratios $4:14:9:9$ as represented in Fig.~2~\cite{strasba1,strasba3}.
This result can readily be calculated using the 6-$j$ formula 
Eq.~(\ref{eqn8}) and some standard orthogonality relations for the 6-$j$ 
symbols. The corresponding calculation in the covariant approach involves 
considerably more labour. Also, the result ``$4+14=9+9$'' for doublet to 
doublet one-pion transitions is a general result which again can easily 
be derived using the 6-$j$ approach\cite{strasba1}.

\newpage

\section{$SU(2N_f)\otimes O(3)$ Structure of the Light-Side 
  Transitions}

Interest in the constituent quark model has recently been rekindled by the 
discovery (or rediscovery) that two-body spin-spin interactions between 
quarks are non-leading in $1/N_C$, at least in the baryon 
sector~\cite{strasba4}. Thus, to leading order in $1/N_C$, light quarks 
behave as if they were heavy and a classification of a light quark system 
in terms of $SU(2N_f)\otimes O(3)$ symmetry multiplets makes sense. 
Transitions between light quark systems are parametrized in terms of a set 
of one-body operators whose matrix elements are then evaluated between the 
$SU(2N_f)\otimes O(3)$ multiplets.

Let us discuss the necessary steps for the implementation of the light-side 
$SU(2N_f)\otimes O(3)$ symmetry in the current transition case. The relevant 
one-body operators are given by~\cite{strasba5}
\medskip\\
{\it $s$-wave to $s$-wave:}
\begin{equation}\label{eqn9}
O=A(\omega)\cdot\one\otimes\one
\end{equation}
{\it $s$-wave to $p$-wave:\ }($(l_K=1;l_k=0)$, $(l_K=0;l_k=1)$ resp.)
\begin{eqnarray}
O_K^\alpha&=&A_K(\omega)v_1^\alpha\one\otimes\one
  +B_K(\omega)(\one\otimes\gamma^\alpha+\gamma^\alpha\otimes\one)\nonumber\\
  O_k^\alpha&=&A_k(\omega)v_1^\alpha\one\otimes\one
  +B_k(\omega)(\one\otimes\gamma^\alpha-\gamma^\alpha\otimes\one).
  \label{eqn10}
\end{eqnarray}
Operators of the type $\gamma^\mu\otimes\gamma_\mu$ or
$v_1^\alpha\gamma^\mu\otimes\gamma_\mu$ are two-body operators and will 
therefore not be included in our discussion. The reduced form factors in 
Eqs.~(\ref{eqn9}) and~(\ref{eqn10}) depend on the velocity transfer 
variable $\omega$ and are unknown functions except for the normalization 
condition $A(1)=1$ in Eq.~(\ref{eqn9}). The operators $\one\otimes\one$ 
and $v_1^\alpha\one\otimes\one$ do not couple angular and spin degrees of 
freedom. These were the operators used some thirty years ago when the 
consequences of the collinear symmetry $SU(6)_W$ or $\tilde U(12)$ were 
worked out. In the following we shall therefore refer to the $A$-type 
operators as the collinear operators. The operators 
$(\one\otimes\gamma^\alpha\pm\gamma^\alpha\otimes\one)$ on the other hand 
introduce spin-orbit coupling interactions and are called spin-orbit 
operators.

The matrix elements of the one-body operators in Eqs.~(\ref{eqn9}) 
and~(\ref{eqn10}) can be readily evaluated using the light-side spin wave 
functions in Table~1. For the $s$-wave to $s$-wave transition the relevant 
matrix element is given by
\begin{equation}\label{eqn11}
A(\omega)\hat{\bar\phi}_{\alpha\beta}^{\nu_1\cdots\nu_{j_2}}
  \hat\phi_{\alpha\beta}^{\mu_1\cdots\mu_{j_1}}.
\end{equation}
There are altogether three ground state to ground state form factors or 
Isgur-Wise functions, one for the $\Lambda_b\rightarrow\Lambda_c$ 
transition and two for the $\{\Sigma_b\}\rightarrow\{\Sigma_c\}$ 
transitions. Equation~(\ref{eqn11}) tells us that they can all be expressed 
in terms of the single form factor $A(\omega)$, where $A(1)=1$ at zero 
recoil. In fact the current transition amplitudes are given 
by~\cite{strasba1,strasba5,strasba6}
\begin{eqnarray}
\Lambda_b\rightarrow\Lambda_c
  &:&M^\lambda=\bar u_2\Gamma^\lambda u_1\frac{\omega+1}2A(\omega)\\
  \{\Sigma_b\}\rightarrow\{\Sigma_c\}
  &:&M^\lambda=\bar\psi_2^\nu\Gamma^\lambda\psi_1^\mu
  (-\frac{\omega+1}2g_{\mu\nu}+\frac12v_{1\nu}v_{2\mu})A(\omega)\nonumber
\end{eqnarray}
where $\Gamma^\lambda=\gamma^\lambda(1-\gamma_5)$ in the Standard Model.
The same result has been obtained by C.K.Chow by analyzing the large
$N_C$ limit of QCD~\cite{strasba7}.
  
When doing a partial wave analysis, the $\Sigma$-type transitions can be 
seen to result from pure $L=0$ diquark transitions. This is a testable 
prediction in as much as the population of the helicity states in the decay 
baryon is fixed resulting in a characteristic angular decay pattern of the 
subsequent decays. More difficult is a test of the relation between the 
$\Lambda$-type and the $\Sigma$-type form factors. In the test one would 
have to compare $\Lambda_b\rightarrow\Lambda_c$ and 
$\Omega_b\rightarrow\{\Omega_c,\Omega_c^*\}$ transitions (where there are 
additional $SU(3)$ breaking effects), since these are the transitions that 
are experimentally accessible. The 
$\Sigma_b\rightarrow\{\Sigma_c,\Sigma_c^*\}$ branching fraction is expected 
to be too small to be measurable.

For the transitions to the $p$-wave charm baryon states one similarly 
reduces the number of reduced form factors when invoking
$SU(2N_f)\otimes O(3)$ symmetry in addition to HQS. For the transition into 
the $K$-multiplet one has a reduction from five HQS reduced form factors 
to the two form factors $A_K(\omega)$ and $B_K(\omega)$ in Eq.~(\ref{eqn10}), 
whereas for transitions into the $k$-multiplet one can relate two HQS 
reduced form factors to the one spin-orbit form factor 
$B_k(\omega)$~\cite{strasba5}. The one-pion and photon transitions can be 
treated in a similar manner. Again one finds a significant simplification 
of the HQS structure, i.e. the number of coupling factors is reduced from 
those listed in Eqs.~(\ref{eqn6}) and~(\ref{eqn7}) when 
$SU(2N_f)\otimes O(3)$ is invoked in addition to HQS. Results for the 
one-pion transitions can be found in~\cite{strasba8}. Corresponding results 
for the photon transitions are presently worked out.

\section{Concluding Remarks}

We have studied the consequences of Heavy Quark Symmetry for current, pion 
and photon transitions between heavy baryons. For the three types of 
transitions we discussed how the most general Heavy Quark Symmetry 
structure can be further simplified by invoking a constituent quark model 
$SU(2N_f)\otimes O(3)$ symmetry for the light-side transition. All of these 
predictions lead to testable results for rates and angular decay 
distributions. The future will show how well these predictions work.

\newpage

\begin{table}
\begin{center}
\begin{tabular}{|c|cccc|}
\hline
& \begin{tabular}{c}light side s.w.f.\\${\hat\phi}^{\mu_1\cdots\mu_j}$
  \end{tabular}
& $j^P$
& \begin{tabular}{c}heavy side s.w.f.\\$\psi_{\mu_1\cdots\mu_j}$
  \end{tabular}
& $J^P$
\\ \hline\hline
\multicolumn{5}{|l|}{$s$-wave states ($l_k=0$, $l_K=0$)}\\
  $\Lambda_Q$&$\hat\chi$&$0^+$&$u$&$\frac12^+$
\\ \hline
  $\{\Sigma_Q\}$
  & ${\hat\chi}^{1\mu_1}$&$1^+$
  & $\begin{array}{c}\frac1{\sqrt3}\gamma^\perp_{\mu_1}\gamma_5u\\
      u_{\mu_1}\end{array}$
  & $\begin{array}{c}\frac12^+\\\frac32^+\end{array}$
\\ \hline\hline
\multicolumn{5}{|l|}{$p$-wave states ($l_k=0$, $l_K=1$)}\\
  $\{\Lambda_{QK1}^{**}\}$
  & ${\hat\chi}^0K_\perp^{\mu_1}$&$1^-$
  & $\begin{array}{c}\frac1{\sqrt3}\gamma^\perp_{\mu_1}\gamma_5u\\
      u_{\mu_1}\end{array}$
  & $\begin{array}{c}\frac12^-\\\frac32^-\end{array}$ 
\\ \hline
  $\Sigma_{QK0}^{**}$
  & $\frac1{\sqrt3}{\hat\chi}^1\cdot K_\perp$&$0^-$&$u$&$\frac12^-$
\\ \hline
  $\{\Sigma_{QK1}^{**}\}$
  & $\frac i{\sqrt2}\varepsilon(\mu_1{\hat\chi}^1K_\perp v)$&$1^{-}$
  & $\begin{array}{c}\frac1{\sqrt3}\gamma^\perp_{\mu_1}\gamma_5u\\
      u_{\mu_1}\end{array}$
  & $\begin{array}{c}\frac12^-\\\frac32^-\end{array}$
\\ \hline
  $\{\Sigma_{QK2}^{**}\}$&
  $\frac12\{{\hat\chi}^{1,\mu_1}K_\perp^{\mu_2}\}_{0}$&$2^-$
  & $\begin{array}{c}\frac1{\sqrt{10}}\gamma_5\gamma^\perp_{\{\mu_1}
      u_{\mu_2\}_0}^{\phantom{\perp}}\\u_{\mu_1\mu_2}\end{array}$
  & $\begin{array}{c}\frac32^-\\\frac52^-\end{array}$
\\ \hline\hline
\multicolumn{5}{|l|}{$p$-wave states ($l_k=1$, $l_K=0$)}\\
  $\{\Sigma_{Qk1}^{**}\}$&${\hat\chi}^0k^{\mu_1}_\perp$&$1^-$
  & $\begin{array}{c}\frac1{\sqrt3}\gamma^\perp_{\mu_1}\gamma_5u\\
      u_{\mu_{1}}\end{array}$
  & $\begin{array}{c}\frac12^-\\\frac32^-\end{array}$
\\ \hline
  $\Lambda_{Qk0}^{**}$&
  $\frac1{\sqrt3}{\hat\chi}^1\cdot k_\perp$&$0^-$&$u$&$\frac12^-$
\\ \hline
  $\{\Lambda_{Qk1}^{**}\}$
  & $\frac i{\sqrt2}\varepsilon(\mu_1{\hat\chi}^1k_\perp v)$&$1^-$
  & $\begin{array}{c}\frac1{\sqrt3}\gamma^\perp_{\mu_1}\gamma_5u\\
      u_{\mu_{1}}\end{array}$
  & $\begin{array}{c}\frac12^-\\\frac32^-\end{array}$
\\ \hline
  $\{\Lambda_{Qk2}^{**}\}$
  & $\frac12\{{\hat\chi}^{1,\mu_1}k_\perp^{\mu_2}\}_0$&$2^-$
  & $\begin{array}{c}\frac1{\sqrt{10}}\gamma_5\gamma^\perp_{\{\mu_1}
      u_{\mu_2\}_0}^{\phantom{\perp}}\\u_{\mu_1\mu_2}\end{array}$
  & $\begin{array}{c}\frac32^-\\\frac52^-\end{array}$
\\ \hline
\end{tabular}
\end{center}
\bigskip
\centerline{\Large\bf Table 1}
\end{table}

\newpage

\centerline{\Large\bf Table Captions}
\vspace{.5cm}
\begin{list}{\bf\rm Tab.~1: }{
\labelwidth1.6cm \leftmargin2.5cm \labelsep0.4cm \itemsep0ex plus0.2ex }
\item Spin wave functions (s.w.f.) of heavy $\Lambda$-type 
  and $\Sigma$-type $s$- and $p$-wave heavy baryons
\end{list}

\centerline{\Large\bf Figure Captions}
\vspace{.5cm}
\newcounter{fig}
\begin{list}{\bf\rm Fig.\ \arabic{fig}: }{\usecounter{fig}
\labelwidth1.6cm \leftmargin2.5cm \labelsep0.4cm \itemsep0ex plus0.2ex }
\item Generic picture of bottom to charm current transitions, and 
  pion and photon transitions in the charm sector in the HQS limit 
  $m_Q\rightarrow\infty$
\item One-pion transition strengths for the transitions 
  $\{\Lambda_{QK2}^{**}\}\rightarrow\{\Sigma_Q\}+\pi$. 
  Degeneracy levels are split for illustrative purposes
\end{list}


\begin{thebibliography}{99}
\bibitem{strasba1}J.G.~K\"orner, M.~Kr\"amer and D.~Pirjol,\\
  Progr.~Part.~Nucl.~Phys.\ {\bf 33} (1994) 787
\bibitem{strasba2}K.~Zalewski, Proceedings of the ``International 
  Europhysics Conference on High Energy Physics'', Marseille 1993, p.~869
\bibitem{strasba3}N.~Isgur and M.B.~Wise,
  Phys.~Rev.~Lett.\ {\bf 66} (1991) 1130
\bibitem{strasba4}E.~Witten, Nucl.~Phys.\ {\bf B223} (1983) 483;\\
  C.~Carone, H.~Georgi and S.~Osofski,
  Phys.~Lett.\ {\bf B322} (1994) 483;\\
  M.~Luty and J.~March-Russel, Nucl.~Phys.\ {\bf B246} (1994) 71;\\
  R.F.~Dashen, E.~Jenkins and A.V.~Manohar,
  Phys.~Rev.\ {\bf D49} (1994) 4713;\\
  R.F.~Dashen, E.~Jenkins and A.V.~Manohar,
  Phys.~Rev.\ {\bf D51} (1995) 3697
\bibitem{strasba5}F.~Hussain, J.G.~K\"orner, J.~Landgraf and S.~Tawfiq,\\
  to be published in Z.~Phys.~C, hep-ph/9505335
\bibitem{strasba6}F.~Hussain, J.G.~K\"orner, M.~Kr\"amer and G.~Thompson,
  Z.~Phys.\ {\bf C51} (1991) 321
\bibitem{strasba7}C.K.~Chow, Phys.~Rev.\ {\bf D51} (1995) 1224;\\
  C.K.~Chow, Cornell Preprint CLNS 95/1392, hep-ph/9601248
\bibitem{strasba8}F.~Hussain, J.G.~K\"orner and S.~Tawfiq,
  to be published
\end{thebibliography}
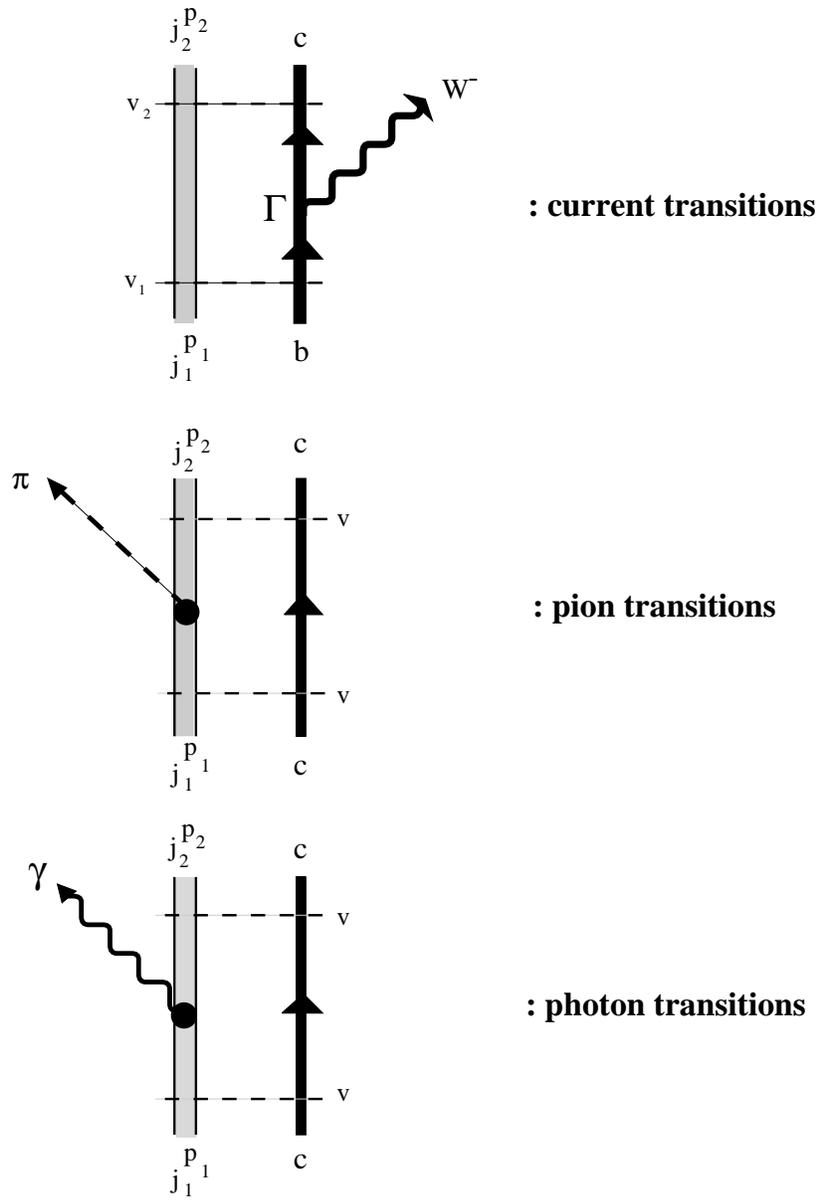
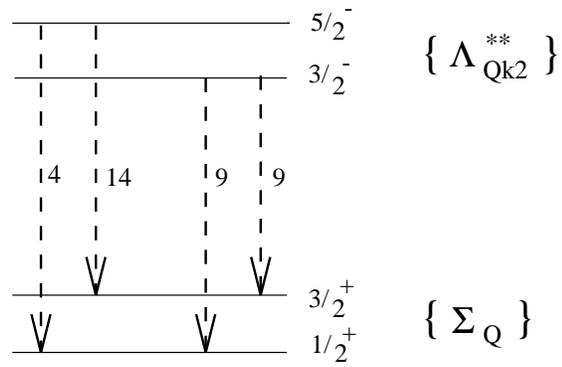
\end{document}